\documentclass{article}
\usepackage[utf8]{inputenc}
\usepackage{amsmath}
\usepackage{graphicx}
\usepackage{amssymb}
\usepackage{cite}
\usepackage{bm}
\usepackage{amsthm}
\usepackage{subcaption}
\usepackage{hyperref}
\usepackage{authblk}
\usepackage{xcolor}
\usepackage[margin=1in]{geometry}

\title{Squared Quartic Hilltop Inflation}

\author{Josh Hoffmann\thanks{Corresponding author: j.hoffmann1@lancaster.ac.uk} } 
\author{David Sloan}
\affil{Department of Physics, Lancaster University, Lancaster UK}

\date{}

\begin{document}

\maketitle
 
\begin{abstract}
    
     A correction to the Quartic Hilltop inflationary model is proposed to account for stabilising terms in the potential. We derive analytical predictions for the spectral index $n_s$ and tensor-scalar ratio $r$ which lie within the Planck 2018 survey bounds. The reheating predictions of the corrected model are investigated and by considering the reheating temperature we further constrain the $(n_s,r)$ parameter space. It is shown that the correction terms are physically important during inflation and generally need to be accounted for when considering Hilltop models as a candidate for inflation.
\end{abstract}

\section{Quartic Hilltop Inflation} 
With the release of the Planck 2018 data it has been possible to accurately test a variety of candidate inflationary models \cite{Akrami:2018odb}. Among the models that best fit the data is the Quartic Hilltop (QH) model. The QH model belongs to the wider class of Hilltop models which are defined by potentials of the form \cite{Boubekeur:2005zm}
\begin{equation}
    V(\phi) = \Lambda\left[1-\lambda\left(\frac{\phi}{m_{\rm pl}}\right)^q +...\right]
\end{equation}
where ... indicates higher order terms. Hilltop models are characterised by inflation occurring near the maxima of the potential on a broad, flat plateau or ``hilltop''. This feature makes Hilltop models attractive for slow roll inflation as the conditions are easy for such a model to fulfil. Furthermore, simple potentials such as these are easier to understand from a particle physics and QFT perspective and do occur in symmetry breaking theories \cite{Kinney:1995ki}, supergravity \cite{PhysRevLett.65.3233,Adams:1992bn,PhysRevD.48.946,Kinney:1995xv,Kumekawa:1994gx,Adams:1996yd,Izawa:1996dv,Izawa:1998rh,Buchmuller:2004tm}, supersymmetry \cite{Covi:2000gx,PhysRevD.65.103518} and superstring theory models \cite{Binetruy:1986ss}. Broadly speaking, inflation models can be divided into two categories \cite{Lyth2007}. These are large field models, where inflation starts with the field at a large value $\phi > m_{\rm pl}$, and small field models for which the inflation starts with the field at a small value $\phi < m_{\rm pl}$.

Hilltop potentials are small field models. These are a widely studied class of models that are easily viewed from the perspective of an effective field theory in particle physics, particularly because the field varies on a scale less than $m_{\rm pl}$, which should be the cutoff of such a theory. They are also good at accommodating slow-roll inflation \cite{Lyth2007,Liddle:1994dx, PhysRevD.29.2162, LINDE1983177}, which provides for excellent agreement with observations. In slow roll inflation, the potential energy of the inflaton field $V(\phi)$ dominates over its kinetic energy and the field evolves slowly. To provide a sufficient amount of slow-roll inflation, potentials which are close to being flat are advantageous. This is well accommodated by small and large field models, whose potentials can be constructed to feature long, flat plateaus for the field to evolve through whilst the universe expands greatly.

The most common matter sources for inflationary theories are constructed from a matter Lagrangian describing usually one (but possibly many) scalar field(s) minimally coupled to gravity \cite{Liddle:1993fq}
\begin{equation}
    S = \int d^4x \sqrt{-g} \left(\frac{1}{2}R + \mathcal{L}_m(\phi, \partial_\mu\phi)\right)
\end{equation}

in units where $8\pi Gc^{-4}=1$, $g_{\mu\nu}$ is the FRLW metric with scale factor $a(t)$ and the Lagrangian for the scalar field is 
\begin{equation}
    \mathcal{L}_m(\phi,\partial_\mu\phi) = \frac{1}{2}\dot{\phi}^2-V(\phi)
\end{equation}

One need only then specify a potential $V(\phi)$ to determine cosmological quantities relevant to inflation. Such a scalar field has a simple equation of motion that includes a friction term from the Hubble parameter $H = \dot{a}/a$ and a force term from the potential $V'$
\begin{equation}
    \ddot{\phi} + 3H\dot{\phi} + V'(\phi) = 0
\end{equation}
with the energy density and pressure given by 
\begin{equation}
\label{energy}
    \rho  = V(\phi) + \frac{1}{2}\dot{\phi}^2
\end{equation}
\begin{equation}
\label{pressure}
    p = -V(\phi) + \frac{1}{2}\dot{\phi}^2
\end{equation}
Inflation requires that the acceleration of the scale factor $\ddot{a}(t)$, given by 
\begin{equation}
    \frac{\ddot{a}}{a} = -\frac{4\pi G}{3}(\rho+3p)
\end{equation}
is positive. This in turn requires the energy density and pressure to satisfy $\rho + 3p < 0$. Given equations (\ref{energy}) \& (\ref{pressure}) this is possible if $\dot{\phi}^2 < 2V(\phi)$, hence the field changes slowly over the course of inflation. In slow roll inflation, the acceleration of the field is also negligibly small $\ddot{\phi} \simeq 0$. The equation of motion then gives the attractor solution 
\begin{equation}
\label{attractor}
    \dot{\phi} \simeq -\frac{1}{3H}V'(\phi)
\end{equation}
By defining the dimensionless slow roll parameters
\begin{equation}
    \varepsilon = \frac{1}{2}m_{\rm pl}^2\left(\frac{V'}{V}\right)^2
\end{equation}
\begin{equation}
    \eta = m_{\rm pl}^2\frac{V''}{V}
\end{equation}
where $m_{\rm pl}$ is the reduced Planck mass, the conditions for slow roll inflation can be summarised as requiring $\varepsilon < 1$, $|\eta| < 1$ and that the field tends to the attractor solution, equation (\ref{attractor}).
One of the greatest successes of inflation is that it provides a mechanism by which small density perturbations in the early universe can later give rise to structure formation \cite{Peacock:1995qb, Linde:1982uu, Dodelson:2003ip, Mukhanov:2003xr, HAWKING1982295, STAROBINSKY1982175, PhysRevLett.49.1110, PhysRevD.28.679}. The main cosmological observables of interest in models such as those described here derive from fluctuations in the scalar field $\phi$ and tensor field $g_{\mu\nu}$, these are the spectral index $n_s(k)$ and the tensor-scalar ratio $r$ \cite{liddle_lyth_2000,Linde:2007fr}. The spectral index is defined through the power spectrum of scalar fluctuations $\mathcal{P}_\phi$
\begin{equation}
    n_s(k) - 1 = \frac{d \ln{\mathcal{P}_\phi}}{d\ln{k}}
\end{equation}
such that if the spectral index does not vary much with the scale $k$, the spectrum follows a power law $\mathcal{P}_\phi \propto k^{n_s-1}$




Hilltop models thus have many attractive features for an inflationary model. In Hilltop models, when the parameter is large enough, $\lambda \gtrsim 1$ inflation occurs at  a small value of the field $\phi/m_{\rm pl} \ll \lambda$ and thus the higher order terms in the potential are heavily suppressed and do not affect inflationary predictions of the model. For this reason, Hilltop models containing only the lowest order terms have usually been considered in previous studies.

The Quartic Hilltop (QH) model invokes a simple potential consisting of a constant term and a quartic term in the field $\phi$.
\begin{equation}
\label{V_QH}
    V(\phi) = \Lambda\left[1-\lambda\left(\frac{\phi}{m_{\rm pl}}\right)^4\right]
\end{equation}
The QH model has recently received more attention in light of the Planck 2018 data, to which it showed a satisfactory fit under a numerical analysis \cite{Martin:2013tda}, whereas earlier attempts at an analytical treatment had ruled QH out on the basis that the predictions for the spectral index $n_s$ as a function of the number of remaining e-folds $N$ of inflation after the scale $k_*$ exits the horizon \cite{Lyth:1998xn}
\begin{equation}
\label{QH_ns_Crude}
    n_s = 1 - \frac{3}{N}
\end{equation}
was too small at $N = 50 $ \& $60$ to fit within the bounds of the Planck 2018 data, as this places $n_s$ between 0.94 and 0.95. To align with the Planck data we require $n_s \gtrsim 0.96$. 

\begin{figure}[h]
  \centering
  \includegraphics[scale=0.5]{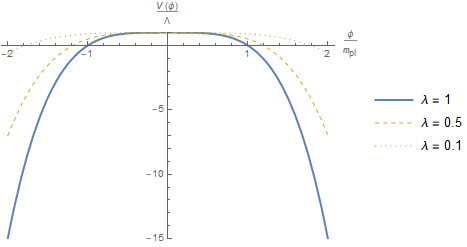}
  \caption{The Quartic Hilltop potential (\ref{V_QH}) of the inflation field $\phi$ at various values of $\lambda$. The numerical values of $\lambda$ chosen for this figure best illustrate the shape of the potential, but the physically interesting values will be for $\lambda \ll 1$.}
  \label{fig:}
\end{figure}

Numerical treatments indicate that the QH model does achieve this, so there is an apparent discrepancy between the results of numerical and analytic investigations. The reason is that in the earlier attempts at solving the model analytically, it was assumed that the physically relevant parameter space has $\lambda \sim 1$. In \cite{Dimopoulos:2020kol}, it is shown that this is in fact not the case and that the parameter $\lambda$ is actually very small. This allows for a more detailed analytic investigation of the QH model, by revising two key assumptions used in deriving the result (\ref{QH_ns_Crude}).

\begin{enumerate}
    \item The value of the inflation field $\phi(N)$ at a given $N$ is much smaller than its vacuum expectation value (VEV) i.e. $\phi(N) \ll \langle \phi \rangle \sim m_{\rm pl}/\lambda^{1/4}$.
    \item The contribution of the value of the inflation field at the end of inflation $\phi_{\rm end}$ to the calculation of $N(\phi)$ is negligible.
\end{enumerate}

By relaxing both of these assumptions it is possible to derive a theoretical prediction for the $r-n_s$ curve which is in close agreement with numerical analysis and thus better fits the Planck 2018 data. In the small $\lambda$ regime, the higher order terms become of greater importance and as we will show in this work, accounting for such terms leads to distinct results when compared to the QH model in the $(n_s,r)$ parameter space favoured by the Planck 2018 survey. Ultimately what is derived in \cite{Dimopoulos:2020kol} is a relationship between the scalar tensor ratio $r$ and the spectral index $n_s$ at a given number of e-folds before the end of inflation $N$.
\begin{equation}
\label{QH_r}
    r(n_s, N) =  \frac{8}{3}(1-n_s)\left[1-\frac{\sqrt{3[2(1-n_s)N-3]}}{(1-n_s)N}\right]
\end{equation}

Whilst this treatment of the QH models yields a result that is much more satisfactory due to being a better fit to the Planck data, the model is not without its issues as discussed in \cite{Kallosh:2019jnl}. Namely, that the QH potential turns negative shortly after inflation ends and is unbounded from below \cite{Kallosh:2019jnl}. This is clearly unsatisfactory as such a theory, when quantized would posses arbitrarily low energy states. The potential needs to be stabilised by the higher order terms that have been neglected in (\ref{V_QH}). 

We will take the investigation of the QH model in \cite{Dimopoulos:2020kol} as a guide for investigating a model which accounts for the higher order terms. That is, we will follow steps analogous to those for investigating the QH model
\begin{itemize}
    \item Compute the number of e-folds, $N(\phi)$, of inflation after the cosmological scale leaves the horizon 
    \item Derive the contribution to $N$ from the inflaton at the end of slow-roll in terms of $\lambda$, $N_{\rm end}(\lambda)$ by solving the conditions $\varepsilon(\phi_{\rm end}) = 1$ \& $|\eta(\phi_{\rm end})| = 1$
    \item Express the spectral index $n_s$ in terms of $N$ and $\lambda$ through our solution for $N_{\rm end}(\lambda)$ 
    \item Invert the expression for $n_s(\lambda, N)$ to find $\lambda(n_s, N)$ and use this to derive the relationship between $r$ and $n_s$ for $N=50$ and $N=60$
\end{itemize}
Although the general method for the QH model and the model containing higher order terms will be the same, a major difference comes in the final step when deriving the relationship between the spectral index and tensor-scalar ratio in which we find that, unlike in the QH case, we find a one-to-many mapping of $n_s$ to $r$ for given values of $N$.

Whilst we will derive analytical results for the relationship between the tensor-scalar ratio and spectral index at a given number of remaining efolds of inflation $N$ which allows accurate comparison of the predictions of the two models. However, this does not necessarily allow us to evaluate how well the Quartic Hilltop model or its corrected version explain the data collected by Planck and other such surveys of the CMB. The $r-n_s$ relationship is only really one half of the picture since, given some model defined by a potential $V(\phi)$ as we are considering here, one may fix any arbitrary $(n_s,r)$ by simply choosing an appropriate time at which the scale $k$ exits the horizon, which translates into an amount of remaining efolds $N$. There are of course some a priori restrictions on $N$ as it is well established that approximately 50 efolds of inflation are required to solve the horizon problem. However as we show in section 4, by considering the reheating dynamics of the corrected model, which one cannot do for the UV-incomplete QH model, it is possible to place further restrictions on the possible values of $N$ so that the resulting temperature of the universe at the end of reheating is consistent with standard cosmology, which may not be the case if one simply chooses $N$ to get a desired $(n_s,r)$ pair.

\section{The Quartic Hilltop Squared Model}

As noted in \cite{Dimopoulos:2020kol}, the Quartic Hilltop (QH) model\footnote{The parameter in the QH model is taken to be $2\lambda$ to keep the form consistent with the series expansion of (\ref{V_QHS}).}
\begin{equation}
    V(\phi) = \Lambda \left[1-2\lambda\left(\frac{\phi}{m_{\rm pl}}\right)^4\right]
\end{equation}

approximates a linear potential near $V(\phi) = 0$ which will become negative for $\phi >  m_{\rm pl}$ shortly before the end of inflation, and needs to be stabilised by higher order terms. This can be achieved by working with the Quartic Hilltop Squared Model (QHS) as suggested in \cite{Kallosh:2019jnl} for which $V(\phi) \geq 0 $.
\begin{equation}
\label{V_QHS}
    V(\phi) = \Lambda \left[1-\lambda\left(\frac{\phi}{m_{\rm pl}}\right)^4\right]^2
\end{equation}

The approach of squaring the Hilltop potential to investigate the effects of accounting for stabilising terms has previously been investigated in the case of the Quadratic Hilltop potential\cite{Martin:2013tda} (its corrected form known as Double Well Inflation) and in \cite{Chowdhury:2019otk} where the authors discuss different aspects of the same corrected Quartic Hilltop potential. As such, it is known that such models, whilst agreeing closely with the uncorrected versions at small field values as one would expect from Taylor expanding the potential, still produce substantially different predictions for the tensor-scalar ratio and spectral index when the VEV is in the super-Planckian regime. Hilltop models generally require a super-Planckian VEV for their predictions of $r$ and $n_s$ to be in lie with current measurements and this is shown for both the Quadratic \cite{Martin:2013tda} and Quartic Hilltop models \cite{Dimopoulos:2020kol}. 

\begin{figure}[h]
  \centering
  \includegraphics[scale=0.6]{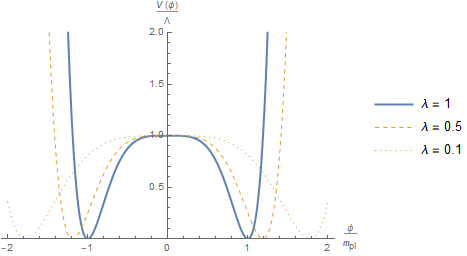}
  \caption{The Quartic Hilltop-Squared potential (\ref{V_QHS}) at various values of $\lambda$. The values of $\lambda$ are chosen to illustrate the shape of the potential, which unlike the QH potential, is bounded from below and gains a local minima from the higher order terms but retains a flat plateau needed for slow-roll inflation.}
  \label{fig:}
\end{figure}

We will calculate analytically the spectral index $n_s$ and tensor-scalar ratio $r$. The analytic calculations can be compared to the detailed data collected by the Planck satellite. Just as the QH model is investigated analytically in \cite{Dimopoulos:2020kol}, analogous steps can be performed for the QHS model. The spectral index and tensor-scalar ratio can be written in terms of the slow roll parameters $\varepsilon$ \& $\eta$.

\begin{equation} 
\label{ns_eq}
    n_s = 1 - 6\varepsilon + 2\eta
\end{equation}
\begin{equation}
\label{r_eq}
    r = 16\varepsilon
\end{equation}
where the slow roll parameters are 
\begin{equation}
\label{eps}
\begin{aligned}
    \varepsilon(\phi) & = \frac{1}{2}m_{\rm pl}^2\left(\frac{V'(\phi)}{V(\phi)}\right)^2 \\ 
    & = 32\lambda^2\left(\frac{\phi}{m_{\rm pl}}\right)^6\frac{1}{\left[1-\lambda\left(\frac{\phi}{m_{\rm pl}}\right)^4\right]^2}
\end{aligned}
\end{equation}
\begin{equation}
\begin{aligned}
    \eta(\phi) & = m_{\rm pl}^2\frac{V''(\phi)}{V(\phi)} \\
    & =\varepsilon(\phi) - 24\lambda\left(\frac{\phi}{m_{\rm pl}}\right)^2\frac{1}{\left[1-\lambda\left(\frac{\phi}{m_{\rm pl}}\right)^4\right]}
\end{aligned}
\end{equation}

It is easiest to work with the spectral index and tensor-scalar ratio in terms of a parameter $Z$ which will be defined shortly and $N$, the remaining number of e-folds until the end of inflation after the cosmological scale has exited the horizon. This is calculated as 
\begin{equation}
\label{N}
\begin{aligned}
    N & = \frac{1}{m_{\rm pl}^2}\int_{\phi_{\rm end}}^{\phi(N)}\frac{V}{V'}d\phi \\
    & = \frac{1}{16}\left(\frac{\phi}{m_{\rm pl}}\right)^2 + \frac{1}{16\lambda}\left(\frac{\phi}{m_{\rm pl}}\right)^{-2} - N_{\rm end}
\end{aligned}
\end{equation}
where the value of the scalar field at the end of inflation is $\phi_{\rm end}$ and  
\begin{equation}
\label{N_end}
     N_{\rm end} = \frac{1}{16}\left(\frac{\phi_{\rm end}}{m_{\rm pl}}\right)^2 + \frac{1}{16\lambda}\left(\frac{\phi_{\rm end}}{m_{\rm pl}}\right)^{-2}
\end{equation}

By defining 
\begin{equation}
    \Bar{N} \equiv N + N_{\rm end} = \frac{1}{16}\left(\frac{\phi}{m_{\rm pl}}\right)^2 + \frac{1}{16\lambda}\left(\frac{\phi}{m_{\rm pl}}\right)^{-2}
\end{equation}
and $Z = 64\lambda\bar{N}^2$, equation (\ref{N}) may be solved for $\left(\frac{\phi}{m_{\rm pl}}\right)^2$ as
\begin{equation}
\label{phi_sol}
    \left(\frac{\phi}{m_{\rm pl}}\right)^2 = 8\bar{N}[Z]
\end{equation}
The function $[Z]$ is defined as 
\begin{equation}
    [Z] = 1 - \sqrt{1-\frac{1}{Z}}
\end{equation}
A property of this function that we will make use of later is that it satisfies $[Z]^2 = 2[Z]-1/Z$. The appearance of this function $[Z]$ sets a constraint $Z > 1$. Since the $Z$ parameter is defined as $Z = 64\lambda\bar{N}^2$, this constraint is saying that for a given value of $\lambda$, which is a free parameter in the model, there will always be a minimum amount of inflation given by 
\begin{equation}
\label{NBar_Bound}
    \bar{N} > \frac{1}{8\sqrt{\lambda}}
\end{equation}
In the QH model there is an analogous bound $\bar{N}_{\rm QH} > 1/4\sqrt{\lambda}$, in fact it is shown that in the small $\lambda$ regime $\bar{N}_{\rm QH} \simeq N + 1/4\sqrt{\lambda}$. This suggests that for the QHS model we should expect $\bar{N} \simeq N + 1/8\sqrt{\lambda}$, and in fact we will later show that this is indeed the case.

Using equation (\ref{phi_sol}), the slow roll parameters can be written as
\begin{equation}
    \varepsilon = \frac{4}{\bar{N}}\frac{Z^2[Z]^3}{\left(1-Z[Z]^2\right)^2}
\end{equation}
\begin{equation}
    \eta = \varepsilon - \frac{3}{\bar{N}}\frac{Z[Z]}{1-Z[Z]^2}
\end{equation}

which allows a calculation of the spectral index and tensor-scalar ratio as 
\begin{equation}
\label{ns}
    n_s = 1 - \frac{6}{\bar{N}}\frac{Z[Z]}{1-Z[Z]^2} - \frac{16}{\bar{N}}\frac{Z^2[Z]^3}{\left(1-Z[Z]^2\right)^2}
\end{equation}
\begin{equation}
\label{r_def}
    r = \frac{64}{\bar{N}}\frac{Z^2[Z]^3}{(1-Z[Z]^2)^2}
\end{equation}

Equation (\ref{ns}) can be written as a quadratic in $F(Z) = \sqrt{Z/(Z-1)}$ by virtue of the purely algebraic relations
\begin{equation}
    1-Z[Z]^2 = 2\left(1-Z[Z]\right)
\end{equation}
\begin{equation}
    \left(\frac{Z[Z]}{1-Z[Z]}\right)^2 = \frac{Z}{Z-1}
\end{equation}
Thus we have 
\begin{equation}
    n_s =  1 + \frac{1}{\bar{N}}\sqrt{\frac{Z}{Z-1}} - \frac{4}{\bar{N}}\left(\frac{Z}{Z-1}\right)
\end{equation}
with the solution
\begin{equation}
\label{F(Z)_Solution}
    \sqrt{\frac{Z}{Z-1}} = \frac{1+\sqrt{1+16\bar{N}\tilde{n}_s}}{8}
\end{equation}
where $\tilde{n}_s = 1 - n_s$.

The goal of this investigation is to produce an analytic relationship between the tensor-scalar ratio $r$ and spectral index $n_s$ at a given number of e-folds before the end of inflation $N$ for $\lambda \ll 1$. Recalling the definitions $Z = 64\lambda\bar{N}^2$ and $\bar{N} = N + N_{\rm end}$, to produce such a function $r(n_s, N)$, we are required to eliminate $\lambda$ \& $\bar{N}$ from the equations.

$N_{\rm end}$ is the contribution of the value of the inflation field at the end of inflation to the total number of e-folds, which as noted in \cite{Dimopoulos:2020kol} can be large for the QH model at small $\lambda$. Investigation of this $N_{\rm end}$ term results in the approximation $\bar{N}_{\rm QH} \simeq N + 1/4\sqrt{\lambda}$, so we are motivated to find an analogous approximation for the QHS model.

Thus far we have extracted both the spectral index and tensor-scalar ratio in terms of two newly defined parameters $Z$ and $\bar{N}$. The end goal of this analysis is to produce a closed form expression for the tensor-scalar ratio in terms of only the spectral index $n_s$ and remaining number of e-folds $N$. A strategy for proceeding can the be devised as follows: An explicit expression for $N_{\rm end}(\lambda)$ will be derived in the small $\lambda$ limit. This allows for the total number of e-folds $\bar{N}$ to be written only in terms of $N$ and $\lambda$ and hence also the $Z$ parameter and spectral index. The resulting expression for $n_s(N,\lambda)$ is invertible for $\lambda(N,n_s)$. The expressions for $\bar{N}(N,\lambda)$ and $\lambda(N,n_s)$ can then be substituted into the definition of $Z$ which is used in equation (\ref{r_def}) to find the target closed form relationship between $r$ and $n_s$ for a given $N$.

\section{Small $\lambda$ behaviour of $N_{\rm end}$}
Guided by the QH model and Planck data, we are mainly interested in the region of parameter space where $\lambda$ is very small, roughly $\lambda \lesssim 10^{-4}$. The small $\lambda$ behaviour of $N_{\rm end}$ can be investigated by considering separately the cases where inflation is ended by the slow roll parameters $\varepsilon$ and $\eta$.

\subsection{$\varepsilon(\phi_{\rm end}) = 1$}
Consider first the case where slow roll inflation is ended by $\varepsilon$ reaching unity before $\eta$. In this case equation (\ref{eps}) becomes
\begin{equation}
\label{phi_end_eqn}
    \frac{1}{\lambda} = \left(\frac{\phi_{\rm end}}{m_{\rm pl}}\right)^3\left(\frac{\phi_{\rm end}}{m_{\rm pl}} + 4\sqrt{2}\right)
\end{equation}
For $\lambda \ll 1$, we have $\phi_{\rm end} \gg m_{\rm pl}$ and thus one has approximately
\begin{equation}
\label{phi_end_eqn_first}
    \left(\frac{\phi_{\rm end}}{m_{\rm pl}}\right)^2 \simeq \frac{1}{\sqrt{\lambda}}
\end{equation}

This result may be substituted into equation (\ref{N_end}) to get an approximate expression for $N_{\rm end}$ in the $\lambda \ll 1$ limit. 
\begin{equation}
    N_{\rm end} \simeq \frac{1}{8\sqrt{\lambda}}
\end{equation}

\subsection{$|\eta(\phi_{\rm end})| = 1$}
Consider now the second case where the end of slow roll is characterized by $|\eta(\phi_{\rm end})|=1$, which of course has two sub cases $\eta(\phi_{\rm end}) = 1$ \& $\eta(\phi_{\rm end}) = -1$.
starting with the $\eta(\phi_{\rm end}) = 1$ case we have
\begin{equation}
\label{x_end+}
    \left(\frac{\phi_{\rm end}}{m_{\rm pl}}\right)^8 - 56\left(\frac{\phi_{\rm end}}{m_{\rm pl}}\right)^6 - \frac{2}{\lambda}\left(\frac{\phi_{\rm end}}{m_{\rm pl}}\right)^4 + \frac{24}{\lambda}\left(\frac{\phi_{\rm end}}{m_{\rm pl}}\right)^2 + \frac{1}{\lambda^2} = 0
\end{equation}
Again, for $\lambda \ll 1$, $\left(\phi_{\rm end}/m_{\rm pl}\right) \gg 1$ and (\ref{x_end+}) is approximately
\begin{equation}
    \left(\frac{\phi_{\rm end}}{m_{\rm pl}}\right)^8 + \frac{1}{\lambda^2} \simeq 0
\end{equation} 
which has no real solutions.

In second sub case $\eta(\phi_{\rm end}) = -1$ we come to the same conclusion. For $\lambda \ll 1$, there are no real, positive solutions. From these cases we conclude that, to a first approximation, for small $\lambda$ the end of slow-roll inflation  is marked only by $\varepsilon(\phi) \geq 1$ .

\subsection{Small $\lambda$ Relationship Between $r$ and $n_s$}
Given the results of section 2, for small $\lambda$ we can use the approximation 
\begin{equation}
    \bar{N} \simeq N + \frac{1}{8\sqrt{\lambda}}
\end{equation}
which should come as no surprise given the earlier bound derived on $\bar{N}$ in equation (\ref{NBar_Bound}). We see that just as in the case of the QH model, the term contributed by $\phi_{\rm end}$ is proportional to $\lambda^{-\frac{1}{2}}$ so this term certainly cannot be neglected when $\lambda \ll 1$.
With this expression we can now write equation (\ref{F(Z)_Solution}) in terms of $\lambda, n_s$ and $N$.

\begin{equation}
    \sqrt{1 + \frac{1}{16N\sqrt{\lambda} + 64N^2\lambda}} = \frac{1}{8}\left[1+\sqrt{1+16\tilde{n}_s\left(N + \frac{1}{8\sqrt{\lambda}}\right)}\right]
\end{equation}

Which can be solved as a quadratic in terms of $\sqrt{\lambda}$, and as such there will be two solutions, both of which are valid. This is of course because $\lambda$ is a free parameter in the model, a given $r$ may map to multiple values of $\lambda$.
\begin{equation}
\label{sqrt_lambda_solution}
    \left(\sqrt{\lambda}\right)_\pm = \frac{12N\tilde{n}_s-2N^2\tilde{n}_s^2-15 \pm \sqrt{8N\tilde{n}_s-15}}{8N\left(15-8N\tilde{n}_s + N^2\tilde{n}_s^2\right)}
\end{equation}

The crossover point between the two branches where $\left(\sqrt{\lambda}\right)_+ = \left(\sqrt{\lambda}\right)_- = \sqrt{\lambda_c}$ is at $\sqrt{8N\tilde{n}_s -15} = 0 \rightarrow N\tilde{n}_s = 15/8$, this also defines the maximum possible value of $n_s$ at a given $N$
\begin{equation}
    n_s^{\rm max} = 1-\frac{15}{8N}
\end{equation}
The calculated spectral index bounds $n_s^{\rm max}(N = 60) = 0.96875$ and $n_s^{\rm max}(N = 50) = 0.9625$ are consistent with the observational data provided by Planck 2018 \cite{Akrami:2018odb}

Substituting $N\tilde{n}_s = \frac{15}{8N}$ into (\ref{sqrt_lambda_solution}) one finds $\sqrt{\lambda}_c = \frac{1}{60N}$, resulting in the critical values of $\lambda_c \simeq 7.72 \times 10^{-8}$ and $\lambda_c \simeq 1.11 \times 10^{-7}$ for $N = 60$ \& $N = 50$ respectively.

The tensor-scalar ratio $r$ which we aim to calculate, can be expressed in terms of the slow-roll parameter $\varepsilon$ as 
\begin{equation}
    \begin{aligned}
        r & = 16\varepsilon \\
        & = \frac{16}{\bar{N}}\left(\frac{Z}{Z-1}\right)\left(1-\sqrt{\frac{Z-1}{Z}}\right)
    \end{aligned}
\end{equation}

Using equation (\ref{F(Z)_Solution}) and the solutions for $\sqrt{\lambda}$ (\ref{sqrt_lambda_solution}), we can calculate the quantities $\bar{N}$ and $Z/(Z-1)$ in terms of $n_s$ and $N$ as desired.
\begin{equation}
\label{r}
    r = \begin{cases}
    \frac{1}{4N}\left(\frac{12N\tilde{n}_s-2N^2\tilde{n}_s^2-15 + \sqrt{8N\tilde{n}_s-15}}{4N\tilde{n}_s-N^2\tilde{n}_s^2+\sqrt{8N\tilde{n}_s-15}}\right)\left[1+g_+(\tilde{n}_s, N)\right]\left[g_+(\tilde{n}_s, N)-7\right], \quad \lambda \geq \lambda_c \\
    \\
    
    \frac{1}{4N}\left(\frac{12N\tilde{n}_s-2N^2\tilde{n}_s^2-15 - \sqrt{8N\tilde{n}_s-15}}{4N\tilde{n}_s-N^2\tilde{n}_s^2 - \sqrt{8N\tilde{n}_s-15}}\right)\left[1+g_-(\tilde{n}_s, N)\right]\left[g_-(\tilde{n}_s, N)-7\right], \quad \lambda < \lambda_c
    \end{cases}
\end{equation}
Where we define the function
\begin{equation}
    g_\pm(\tilde n_s, N) = \sqrt{\frac{12N\tilde n_s+62N^2\tilde n_s^2-16N^3\tilde n_s^3-15\pm(1+16N\tilde n_s)\sqrt{8N\tilde n_s-15}}{12N\tilde n_s-2N^2\tilde n_s^2-15 \pm \sqrt{8N\tilde n_s-15}}}
\end{equation}

\begin{figure}[h]
  \centering
  \includegraphics[scale=0.7]{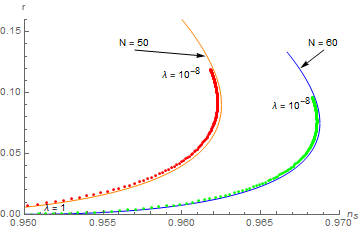}
  \caption{Analytical solutions (bold) for the spectral index $n_s$ and tensor-scalar ratio $r$ of the QHS model equation (\ref{r}), compared with their numerical solutions (dotted) for $10^{-8} \leq \lambda \leq 1$ at $N = 50$ and $N = 60$ e-folds before the end of inflation.}
  \label{fig:QHS_r_ns_N}
\end{figure}
\begin{figure}[h]

\begin{subfigure}[b]{.6\textwidth}
  \centering
  \includegraphics[scale = 0.45]{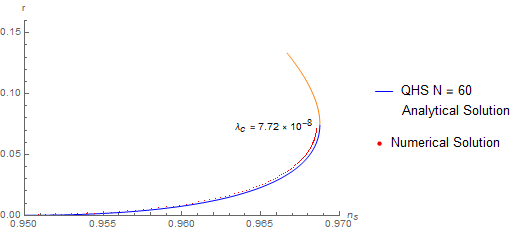}
  \caption{QHS $N = 60$ numerical solution calculated up to the \newline crossover point between $r_\pm$ branches at $\lambda_c$ compared to \newline analytical solution}
  \label{fig:60_c}
\end{subfigure}%
\hfill
\begin{subfigure}[b]{.6\textwidth}
  \centering
  \includegraphics[scale = 0.45]{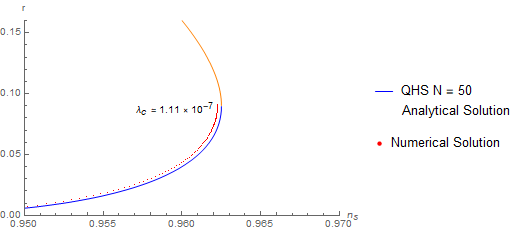}
  \caption{QHS $N = 50$ numerical solution calculated up \newline to the crossover point between $r_\pm$ branches at $\lambda_c$ \newline compared to analytical solution}
  \label{fig:50_c}
\end{subfigure}
\end{figure}

By comparison to numerical solutions for a broad range of $\lambda$ from $\lambda \ll 1$ to $\lambda \approx 1$ in figure (\ref{fig:QHS_r_ns_N}) we can see that this approximation is quite accurate, even in the regime when $\lambda$ is large ($\lambda \sim 1$). We can further validate the analytical solution by comparing the $N = 60$ and $N = 50$ solutions to numerical solutions calculated up to the predicted critical $\lambda_c$ values.
In figures (\ref{fig:60_c}) and (\ref{fig:50_c}) it can be seen that  if the system is solved numerically up to the predicated $\lambda_c$ values, the solution branches do indeed stop at the crossover point between the $r_+$ and $r_-$ branches of the analytical solution.

\begin{figure}[h]
  \centering
  \includegraphics[scale=0.7]{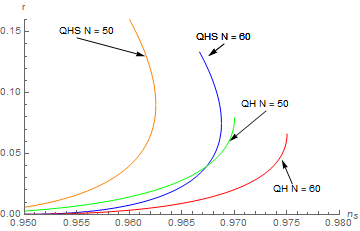}
  \caption{Comparisons of the analytic $r(n_s)$ solutions for Quartic Hilltop-Squared (\ref{r}) calculated in this work and Quartic Hilltop (\ref{QH_r}) calculated in \cite{Dimopoulos:2020kol} at $N = 50$ \& $N = 60$ e-folds before the end of inflation.}
  \label{Model_Comp}
\end{figure} 

Comparing the solution for the QH model (\ref{QH_r}) and QHS model (\ref{r}), we see that the effect of including the higher order stabilising terms causes the predictions of the QHS model to diverge quite dramatically from those of the QH model. The models begin to diverge around $n_s \simeq 0.95$ which is well outside of the region expected from the Planck 2018 data making them quite distinct within the acceptable regions. This is an important result as it makes clear that different methods of stabilising the potential must be considered as distinct if we are to take Hilltop potentials as serious candidates for inflation. It is not sufficient to ignore these stabilising terms in the potential since in the small $\lambda$ regime they have a dramatic effect on inflationary predictions of the theory.

\begin{figure}[h]
  \centering
  \includegraphics[scale=0.6]{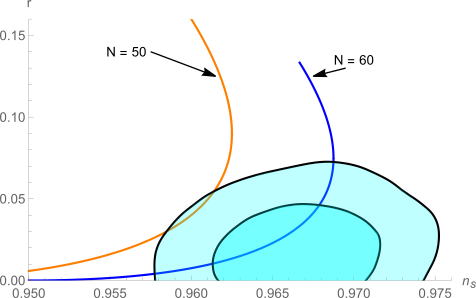}
  \caption{Analytical QHS solutions (bold) given by equation (\ref{r}) at $N = 50$ \& $N = 60$ e-folds before the end of inflation, compared to the Planck 2018 data bounds represented by the filled regions ($2\sigma$ in lighter, $1\sigma$ in darker).}
  \label{QHS_Bounds}
\end{figure} 
The result of including the stabilising terms, as shown in figure (\ref{QHS_Bounds}), is that the N = 50 branch has shifted out of the $1\sigma$ region whilst remaining just within the $2\sigma$ bound. In the QH model, both branches remains within the $1\sigma$ region, however this does not spell such bad news for the QHS model since $N \simeq 60$ is generally favoured by reheating constraints.

Another feature that is seen in the QHS model but not the QH model is that the solution for $r(n_s)$ is not a function, given $n_s$ may correspond to multiples values of $r$. As such, the curves in figure \ref{fig:QHS_r_ns_N} representing the pass vertical and. The points at which the curves reach their maximal values of the spectral index are indeed the bounds calculated earlier $n_s^{\rm max}(N = 60) = 0.96875$ and $n_s^{\rm max}(N = 50) = 0.9625$. This arises due to the stabilisation, and is a general feature of stabilised Hilltop models. This can be seen by considering the spectral index $n_s$ and tensor ratio as functions of the slow roll parameters

\begin{equation}
\begin{aligned}
    n_s & = 1 - 6\varepsilon(\phi) + 2\eta(\phi) \\
    r & = 16\varepsilon(\phi)
\end{aligned}
\end{equation}

In the QHS case, the slow roll parameters are 
\begin{equation}
\label{nsr}
    \eta(\phi) = \varepsilon(\phi) - 24\lambda\left(\frac{\phi}{m_{\rm pl}}\right)^2\frac{1}{1-\lambda\left(\frac{\phi}{m_{\rm pl}}\right)^4}
\end{equation}
Defining 
\begin{equation}
    \Gamma(\phi,\lambda) = 24\lambda\left(\frac{\phi}{m_{\rm pl}}\right)^2\frac{1}{1-\lambda\left(\phi/m_{\rm pl}\right)^4}
\end{equation}
the two equations in (\ref{nsr}) can be combined to give
\begin{equation}
    r = 8\left(1-n_s\right) - 32\Gamma(\phi,\lambda)
\end{equation}
From the analytical results in this work, specifically equation (\ref{sqrt_lambda_solution}), for a given $n_s<n_s^{\rm max}$ there will be one or two corresponding values of lambda $\lambda_\pm$ (one in the case $N\tilde{n}_s = 15/8$ and two otherwise). And thus for a given $n_s$ there will be one or two values of $\Gamma(\phi,\lambda)$ and hence one or two corresponding values of $r$. The QH model does not exhibit this behaviour because in this particular case $\eta(\phi)$ and $\varepsilon(\phi)$ are linearly independent. From a physical standpoint, we expect this to be a generic feature of all Hilltop models that take higher order correction terms into account. Consider the spectral index $n_s$ written in terms of the Hilltop potential $V(\phi)$.

\begin{equation}
    n_s = 1 - 3m_{\rm pl}^2\left(\frac{V'(\phi)}{V(\phi)}\right)^2 + 2m_{\rm pl}^2\frac{V''(\phi)}{V(\phi)}
\end{equation}

The dependence of $n_s$ on the ratios of the derivatives of the potential to the potential itself suggests that we should write this in terms of $W(\phi) = \ln V(\phi)$ for which

\begin{equation}
    n_s = 1 - m_{\rm pl}^2W'(\phi) + 2m_{\rm pl}^2W''(\phi)
\end{equation}

If one includes the stabilising terms in $V(\phi)$ then the potential will have a local minima and thus the second derivative will also have a local minima $\phi_{min} < \phi_{\rm end}$ for sufficiently small $\lambda$. Because the curvature of $V(\phi)$ has a turning point, $W''(\phi)$ will have a form such that if we fix a particular value $W''_*= W(\phi_0,\lambda_0)$, we can find another set $(\phi_1,\lambda_1)$ for which $W''_* = W(\phi_1,\lambda_1)$. Since the spectral index is just a linear combination of the first and second derivatives it will also inherit this property. So for a fixed value of the spectral index $n_s^*$ at $(\phi_0,\lambda_0)$, we may also find another set $(\phi_1,\lambda_1)$ which corresponds to the same value of the spectral index. However the gradient term $W'(\phi)$ is not necessarily the same at $(\phi_0,\lambda_0)$ and $(\phi_1,\lambda_1)$ and since the tensor-scalar ratio is proportional to only this term, $n_s^*$ will necessarily correspond to multiple values of $r$.

\section{Reheating in QHS}
\subsection{Generic Reheating Analysis and Subsequent Bound on $w_{re}$}
Whilst it is useful to be able to calculate $(n_s,r)$ pairs for the Quartic Hilltop Squared model, and verify that for $50 \lesssim N \lesssim 60$ the predictions of the model lie within the parameter space determined by Planck 2018 data, this does not fully validate the model as a viable candidate for inflation. As discussed in section D of \cite{Martin:2010kz}, given a generic potential $V(\phi)$ one may calculate and desired $(n_s,r)$ pair by through equations \ref{ns_eq} \& \ref{r_eq} by choosing an appropriate time at which the scale $k$ exits the horizon. This will correspond to a number of efolds of inflation $N_k$ remaining  when the scale exits. This however, may not result in acceptable reheating predictions of the model defined by $V(\phi)$. The number of reheating efolds and the temperature at the end of reheating $T_{re}$ depends explicitly on $N_k$\cite{Martin:2010kz,Cook:2015vqa}. This dependence is derived by considering the evolution of the horizon scale and the energy density throughout the evolution of the universe.

\begin{figure}[h]
    \centering
    \includegraphics[scale = 0.6]{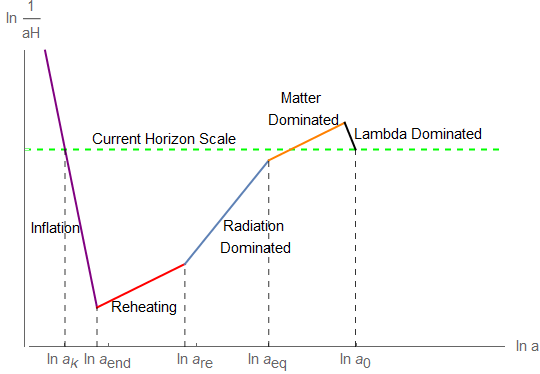}
    \caption{A schematic outlining the evolution of the horizon scale $(aH)^{-1}$ as a function of the (log) scale factor $a(t)$ and dominant matter source at each time from inflation to the present day at $a_0(t)$.}
    \label{HScale}
\end{figure}

Let the comoving scale at the time it exits the horizon be $k = a_kH_k$, this can be compared to the horizon scale today
\begin{equation}
    \frac{k}{a_0H_0} = \frac{a_k}{a_{end}}\frac{a_{end}}{a_{re}}\frac{a_{re}}{a_{eq}}\frac{a_{eq}}{a_0}\frac{H_{eq}}{H_0}\frac{H_k}{H_{eq}}
\end{equation}
We the write $N_k = \ln{a_{end}/a_k}$ for the number of remaining inflation efolds after the scale exits the horizon, $N_{re} = \ln{a_{re}/a_{end}}$ for the number of reheating efolds, $N_{RD} = \ln{a_{eq}/a_{re}}$ for the number of efolds during radiation domination.

\begin{equation}
\label{k/a}
    \ln{\frac{k}{a_0H_0}} = - N_k - N_{re} - N_{RD} - \ln{\frac{a_{eq}}{a_0}} + \ln{\frac{H_{eq}}{H_0}} + \ln{\frac{H_k}{H_{eq}}}
\end{equation}

Assuming a constant equation of state during reheating $w_{re}$, we may track the evolution of the energy density from the end of inflation to the end of reheating

\begin{equation}
    \frac{\rho_{end}}{\rho_{re}} = \left(\frac{a_{end}}{a_{re}}\right)^{-3(1+w_{re})}
\end{equation}

In terms of the reheating efolds this is 
\begin{equation}
    N_{re} = \frac{1}{3(1+w_{re})}\ln{\frac{\rho_{end}}{\rho_{re}}}
\end{equation}

where the energy densities are given by 
\begin{equation}
    \rho_{re} = \frac{\pi^2}{30}g_{re}T_{re}^4
\end{equation}
\begin{equation}
\label{p_re}
    \rho_{end} = \frac{3}{2}V_{end}
\end{equation}
where $V_{end}$ is the value of the potential at the end of inflation (i.e. when $\varepsilon \simeq 1$) , $g_{re}$ is the effective number of relativistic species at the end of reheating, and the reheating temperature $T_{re}$ can be written in terms of the CMB temperature today $T_0$
\begin{equation}
\label{T_0}
    \frac{T_{re}}{T_0} = \left(\frac{43}{11g_{re}}\right)^{\frac{1}{3}}\frac{a_0}{a_{eq}}\frac{a_{eq}}{a_{re}}
\end{equation}
 By taking the logarithm of Eq. \ref{T_0} and combining with Eq. \ref{k/a} we obtain the reheating efolds and temperature (Equivalently one may also consider the reheating energy density \ref{p_re} rather than the temperature as is done in some literature \cite{Martin:2010kz}).

\begin{equation}
    \label{N_re}
     N_{re} = \frac{4}{1-3w_{re}}\left(\frac{1}{4}\ln{\frac{\pi^2g_{re}}{45}} + \frac{1}{3}\ln{\frac{11}{43g_{re}}} + \ln{\frac{a_0T_0}{k}}-\ln{\frac{V_{end}^{\frac{1}{4}}}{H_k}} - N_k\right)
\end{equation}
\begin{equation}
\label{T_re}
  \left[\left(\frac{43}{11g_{re}}\right)^{\frac{1}{3}}\left(\frac{a_0T_0}{k}\right)H_ke^{-N_k}\left[\frac{45V_{end}}{\pi^2g_{re}}\right]^{-\frac{1}{3(1+w_{re})}}\right]^{\frac{3(1+w_{re})}{3w_{re}-1}} 
\end{equation}

For the model to at least be considered as an acceptable candidate one needs both that the predicted $(n_s,r)$ are within the bounds of current measurements and that the corresponding remaining number of inflation efolds $N_k$ gives an acceptable reheating temperature $T_{re}$.
We may calculate analytically the model dependant terms in equations \ref{N_re} \& \ref{T_re}.
Firstly the value of the potential at the end of inflation is simply $V(\phi)$ evaluated on the solution to equation \ref{phi_end_eqn}. However using just the first-order approximate solution \ref{phi_end_eqn_first} would result in $V_{end} = 0$. This may be sufficient for calculating $N_{re}$ since it only depends logarithmically on $V_{end}$, but the reheating temperature $T_{re}$ is sensitive to $V_{end}$ so one needs at least the second order term in the power series solution to \ref{phi_end_eqn} to calculate the reheating temperature.
\begin{equation}
    \left(\frac{\phi_{end}}{m_{pl}}\right) \simeq\lambda^{-\frac{1}{4}} + 3\lambda^{-\frac{7}{4}} \qquad \lambda \ll 1
\end{equation}
and thus we have the value of the potential at the end of inflation 
\begin{equation}
    V_{end} \simeq 144\Lambda\lambda^{-3}
\end{equation}
The value of the Hubble parameter $H_{k}$ at the pivot scale is  
\begin{equation}
    H_k \simeq \pi m_{pl}\sqrt{\frac{A_sr}{2}}
\end{equation}
Finally, given $N_{end}\simeq 1/8\sqrt{\lambda}$ and equation \ref{N}, the number of remaining efolds of inflation $N_{k}$ may be written in terms of the field value at the pivot scale $\phi_k$.
\begin{equation}
    N_{k} = \frac{1}{16}\left(\frac{\phi_k}{m_{pl}}\right)^2+\frac{1}{16\lambda}\left(\frac{\phi_k}{m_{pl}}\right)^{-2} - \frac{1}{8\sqrt{\lambda}}
\end{equation}
And thus the reheating efolds and temperature are expressed as 

\begin{align*}
    N_{re} = \frac{4}{1-3w_{re}} &\left[  \frac{1}{4}\ln{\frac{\pi^2g_{re}}{45}} + \frac{1}{3}\ln{\frac{11}{43g_{re}}} + \ln{\frac{a_0T_0}{k}} -\ln\left[\frac{\sqrt{2}(144\Lambda)^{\frac{1}{4}}}{\lambda^{\frac{3}{4}}\pi m_{pl}\sqrt{A_sr}}\right] \right. \\
    & \left. + \frac{1}{8\sqrt{\lambda}}-\frac{1}{16}\left(\frac{\phi_k}{m_{pl}}\right)^2 - \frac{1}{16\lambda}\left(\frac{\phi_k}{m_{pl}}\right)^{-2} \right]
\end{align*}
\begin{align*}
    T_{re} = \left[\left(\frac{43}{11g_{re}}\right)^{\frac{1}{3}}\left(\frac{a_0T_0}{k}\right)\pi m_{pl}\sqrt{\frac{A_sr}{2}}e^{-N_k}\left[\frac{6480\Lambda}{\lambda^3}\pi^2g_{re}\right]^{-\frac{1}{3(1+w_{re})}}\right]^{\frac{3(1+w_{re})}{3w_{re}-1}}
\end{align*}
To investigate the reheating predictions of the model we consider $r-n_s$ curves for $50 \leq N_k \leq 65$ and their corresponding reheating efolds and temperatures.
\begin{figure}[h]
    \centering
    \includegraphics[scale = 0.6]{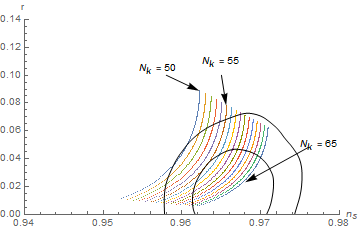}
    \caption{QHS numerical solutions of the tensor-scalar ratio $r$ and spectral index $n_s$ for $50 \leq N_k \leq 65$ over the parameter range $10^{-3}\lesssim \lambda \lesssim 10^{-7}$. The solid black lines represent the Planck 2018 bounds just as in figure \ref{QHS_Bounds} }
    \label{QHS_Reheating}
\end{figure}

\begin{figure}[h]
    \centering
     \begin{subfigure}[b]{0.48\textwidth}
         \centering
         \includegraphics[width=\textwidth]{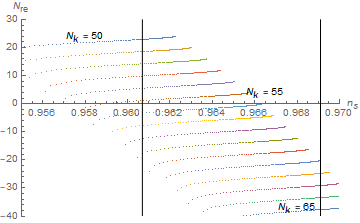}
         \caption{Plots of the reheating efolds $N_{re}$ corresponding to the numerical $(n_s,r)$ solutions for the QHS model in figure \ref{QHS_Reheating} with $w_{re} = 0$. The solid vertical black lines mark the Planck 2018 $1\sigma$ bound on the spectral index $n_s$. For $N_k > 55$ all of the constant $N_k$ curves lie below the $N_{re} = 0$ line, thus in this case the model does not produce any reheating if there are more than 55 efolds of inflation. }
         \label{QHS_Reheating_N_w_0}
     \end{subfigure}
     \hfill
     \begin{subfigure}[b]{0.48\textwidth}
         \centering
         \includegraphics[width=\textwidth]{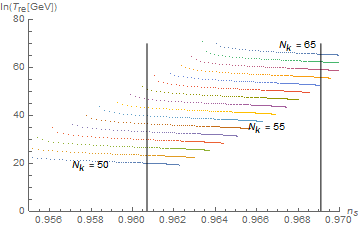}
         \caption{Plots of the (log) reheating temperature $T_{re}$ corresponding to the numerical $(n_s,r)$ solutions for the QHS model in figure \ref{QHS_Reheating} with $w_{re} = 0$. The solid vertical black lines mark the Planck 2018 $1\sigma$ bound on the spectral index $n_s$.}
         \label{QHS_Reheating_T_w_0}
     \end{subfigure}
     \caption{}
     \label{QHS_w_0}
\end{figure}
From the plots of the reheating efolds in figure \ref{QHS_Reheating_N_w_0}, one sees that if there are more than approximately 55 remaining efolds of inflation when the pivot scale exits the horizon, then it it not possible to get $N_{re} > 0$. However for $N_k\leq55$, all of the $(n_s,r)$ predictions of the model lie outside of the $1\sigma$ bound in figure \ref{QHS_Reheating} and thus an equation of state parameter of $w_{re} = 0$ leaves the model with no acceptable predictions compared to the Planck 2018 data.

\begin{figure}[h]
    \centering
     \begin{subfigure}[b]{0.48\textwidth}
         \centering
         \includegraphics[width=\textwidth]{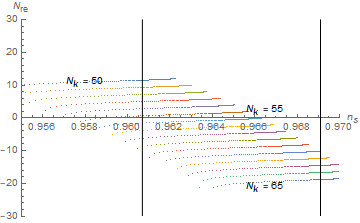}
         \caption{Plots of the reheating efolds $N_{re}$ corresponding to the numerical $(n_s,r)$ solutions for the QHS model in figure \ref{QHS_Reheating} with $w_{re} = -1/3$. The solid vertical black lines mark the Planck 2018 $1\sigma$ bound on the spectral index $n_s$. Just as in figure \ref{QHS_Reheating_N_w_0} The model cannot produce any reheating efolds if there are more than 55 efolds of inflation.}
         \label{QHS_Reheating_N_w_-1/3}
     \end{subfigure}
     \hfill
     \begin{subfigure}[b]{0.48\textwidth}
         \centering
         \includegraphics[width=\textwidth]{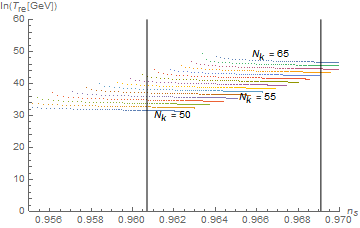}
         \caption{Plots of the (log) reheating temperature $T_{re}$ corresponding to the numerical $(n_s,r)$ solutions for the QHS model in figure \ref{QHS_Reheating} with $w_{re} = -1/3$. The solid vertical black lines mark the Planck 2018 $1\sigma$ bound on the spectral index $n_s$.}
         \label{QHS_Reheating_T_w_-1/3}
     \end{subfigure}
     \caption{}
     \label{QHS_w_minus_1over3}
\end{figure}
\clearpage

\begin{figure}[t]
    \centering
     \begin{subfigure}[b]{0.48\textwidth}
         \centering
         \includegraphics[width=\textwidth]{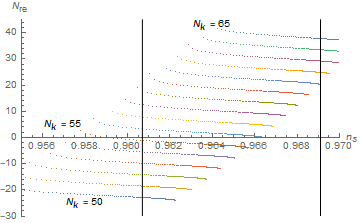}
         \caption{Plots of the reheating efolds $N_{re}$ corresponding to the numerical $(n_s,r)$ solutions for the QHS model in figure \ref{QHS_Reheating} with $w_{re} = 2/3$. The solid vertical black lines mark the Planck 2018 $1\sigma$ bound on the spectral index $n_s$. In this case the behaviour of the reheating efolds is inverted from figures \ref{QHS_Reheating_N_w_0} \& \ref{QHS_Reheating_N_w_-1/3}, the model only produces non-zero reheating efolds $N_{re}\geq 0$ for $N_k \geq 55$. However, also restricting to the region of the spectral index's $1\sigma$ bound makes this inequality noninclusive $N_k > 55$. }
         \label{QHS_Reheating_N_w_2/3}
     \end{subfigure}
     \hfill
     \begin{subfigure}[b]{0.48\textwidth}
         \centering
         \includegraphics[width=\textwidth]{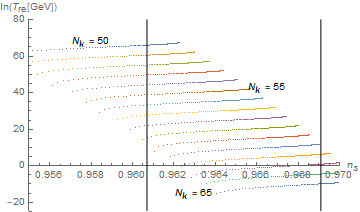}
         \caption{Plots of the (log) reheating temperature $T_{re}$ corresponding to the numerical $(n_s,r)$ solutions for the QHS model in figure \ref{QHS_Reheating} with $w_{re} = 2/3$. The solid vertical black lines mark the Planck 2018 $1\sigma$ bound on the spectral index $n_s$. }
         \label{QHS_Reheating_T_w_2/3}
     \end{subfigure}
     \caption{}
     \label{QHS_w_2over3}
\end{figure}

\begin{figure}[h]
    \centering
     \begin{subfigure}[b]{0.48\textwidth}
         \centering
         \includegraphics[width=\textwidth]{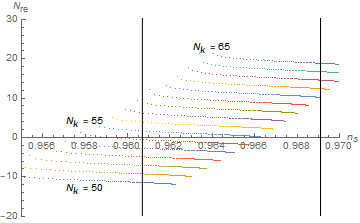}
         \caption{Plots of the reheating efolds $N_{re}$ corresponding to the numerical $(n_s,r)$ solutions for the QHS model in figure \ref{QHS_Reheating} with $w_{re} = 1$. The solid vertical black lines mark the Planck 2018 $1\sigma$ bound on the spectral index $n_s$. As in figure \ref{QHS_Reheating_N_w_2/3}, the constant $N_k$ curves lie above the $N_{re} = 0$ line in the $1\sigma$ $n_s$ region for $N_k > 55$.   }
         \label{QHS_Reheating_N_w_1}
     \end{subfigure}
     \hfill
     \begin{subfigure}[b]{0.48\textwidth}
         \centering
         \includegraphics[width=\textwidth]{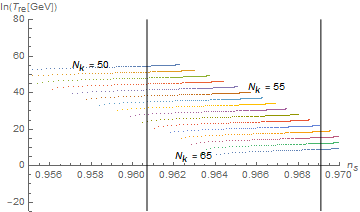}
         \caption{Plots of the (log) reheating temperature $T_{re}$ corresponding to the numerical $(n_s,r)$ solutions for the QHS model in figure \ref{QHS_Reheating} with $w_{re} = 1$. The solid vertical black lines mark the Planck 2018 $1\sigma$ bound on the spectral index $n_s$. }
         \label{QHS_Reheating_T_w_1}
     \end{subfigure}
     \caption{}
     \label{QHS_w_1}
\end{figure}
\clearpage
Taking a negative equation of state parameter $w_{re} = -1/3$ as in figures \ref{QHS_Reheating_N_w_-1/3} and \ref{QHS_Reheating_T_w_-1/3} serves to push the solutions slightly tighter together compared to the $w_{re} = 0$ case, as one would expect given that the dependence of the reheating efolds $N_{re}$ on $w_{re}$ follows $(1-3w_{re})^{-1}$. In this case there are still no acceptable solutions for $N_{k} > 55$. We may however also try positive equation of state parameters.

Using a positive equation of state parameter in figures \ref{QHS_w_2over3} and \ref{QHS_w_1} inverts the behaviour of $N_{re}$ and $T_{re}$ with respect to $N_{k}$ so that the number of reheating efolds increases with $N_k$ and the reheating temperature decreases for with $N_k$. In this case there are now solutions which lie in the $1\sigma$ region of figure \ref{QHS_Reheating} (i.e. $N_k > 55$) and have $N_{re} > 0$.

\subsection{Reheating Temperature Restrictions On The $r-n_s$ Parameter Space.}
In section 4.1 we have established that in order to get a non-zero amount of reheating efolds from the QHS model consistent with Planck measurements of the spectral index $n_s$, one is restricted to a positive equation of state parameter $ 0 < w_{re} \leq 1$. The temperature of the universe at the end of reheating is loosely bounded from below in order to be consistent with standard cosmology. If reheating is to occur before Big Bang Nucleosynethsis (BBN) then one requires $T_{re} > T_{BBN}$. BBN occurs on temperature scales of roughly $T_{BBN} \lesssim 10 \text{MeV}$ \cite{Steigman:2007xt} so the reheating temperature is constrained to be at least larger than $0.01\text{GeV}$. The reheating temperature is also loosely bounded from above by the fact that it should not exceed the energy scale of inflation $T_{re} \lesssim 10^{16} \text{GeV}$ \cite{Liddle:1993ch}, but this bound may be further restricted if one accepts supersymmetry and considers the effects of gravitino production during inflation on BBN \cite{Khlopov:1984pf, Kawasaki:2004yh, Copeland:2005qe, Kallosh:1999jj}. This restricts the reheating temperature to much smaller range $0.01 \text{GeV} \lesssim T_{re} \lesssim 10^8 \text{GeV}$. Considering the extremal case where the equation of state parameter $w_{re} = 1$, the reheating temperature bounds can be used to restrict the QHS model to a much tighter region of the $r-n_s$ parameter space, whose reheating efolds are non-zero and reheating temperature is compatible with standard cosmology.

\begin{figure}[h]
    \centering
    \includegraphics[scale = 0.4]{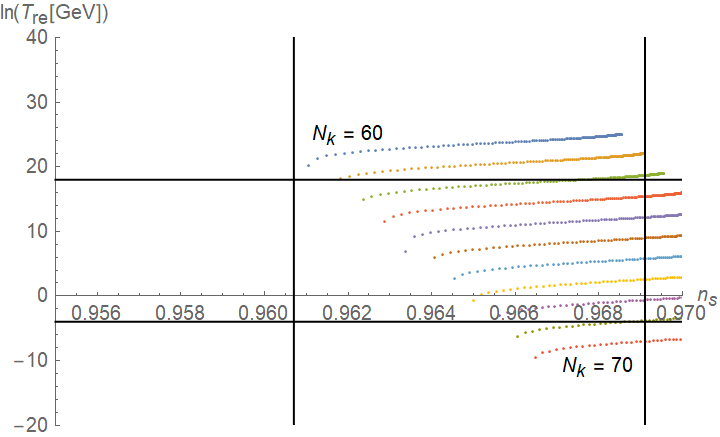}
    \caption{We plot the (log) reheating temperature of the QHS model over the range $60 \leq N_k \leq 70$ where as in previous figures the vertical black lines mark the $1\sigma$ Planck 2018 bounds on the spectral index $n_s$. The horizontal black lines mark the reheating temperature bounds $0.01 \text{GeV} \lesssim T_{re} \lesssim 10^8 \text{GeV}$. The range of the parameter $\lambda$ is the same as in the figure \ref{QHS_Reheating}.}
    \label{QHS_T_Bounds}
\end{figure}
\begin{figure}[h]
    \centering
    \includegraphics[scale = 0.4]{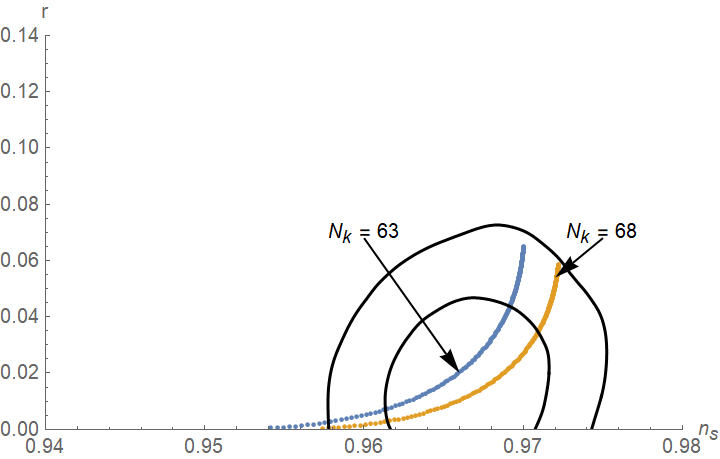}
    \caption{The reheating temperature limits restrict us to the area bounded by the $N_k = 63$ and $N_k = 68$ curves in the $r-n_s$ parameter space which is considerably smaller than just the Planck 2018 bounds.} 
    \label{QHS_Spec_Bounded}
\end{figure}

If we consider the fixed $N_k$ curves in figure \ref{QHS_T_Bounds} who lie within the reheating temperature bounds for all values of $\lambda$ within the $1\sigma$ region, one is restricted to the range $63 \leq N_k \leq 68$ of remaining efolds of inflation. In figure \ref{QHS_Spec_Bounded} we plot the $r-n_s$ curves for $N_k = 63$ and $N_k = 68$. These curves bound a region of parameter space that is both compatible with the Planck 2018 measurements and such that the model produces reheating temperatures within the range $0.01 \text{GeV} \lesssim T_{re} \lesssim 10^8 \text{GeV}$.

\newpage

\section{Discussion}

    We have derived analytical predictions for the cosmological parameters $n_s$ and $r$ in a Quartic Hilltop Squared model. This model accounts for a higher order correction term which stabilises the potential in \cite{Dimopoulos:2020kol}, such that it is nowhere negative. The work of \cite{Dimopoulos:2020kol} makes it clear that when working with Hilltop models in the small $\lambda$ regime (the one in which they best fit observational data), one cannot assume that when calculating the remaining number of e-folds of inflation, the contribution of the inflaton field at the end of inflation will be small enough to be negligible. In fact one find that the contribution is $N_{\rm end} \propto \lambda^{-1/2}$. Clearly this will be large for small $\lambda$. In the case of the Quartic Hilltop model the relevant range of values of $\lambda$ are $\lambda \leq 10^{-4}$, yet one only needs $\lambda = 4 \times 10^{-4}$ for the $\lambda^{-1/2}$ contribution to be as large as 50 e-folds. In this paper we find that the same relationship exists when accounting for the higher order terms in the Quartic Hilltop Squared model. The only difference is that the constant of proportionality is smaller. For the QH model $N_{\rm end}^{\rm QH}/\lambda^{-1/2} = 1/4$ and for the QHS model $N_{\rm end}^{\rm QHS}/\lambda^{-1/2} = 1/8$. Because $N_{\rm end}$ and $\lambda$ follow the same relationship in the QHS model studied in this paper we also conclude that one cannot assume this contribution to be small in the calculation of the remaining number of e-folds and that it must be accounted for in order to explain why the model fits well to observational data.
    
    When the analytical solution derived in this work is compared with that of the QH model in \cite{Dimopoulos:2020kol} and the Planck 2018 data there are two important observations to be made. Firstly, both Hilltop models provide very good fits to the observational data, being within the $1\sigma$ region with the exception only of the QHS $N = 50$ branch. The QH model was already known to be a good candidate inflationary model and we show here that whilst accounting for the higher order terms does shift the positions of the $N$-branches quite dramatically, the model still remains within a region of the parameter space that makes it an attractive inflationary model. In addition to both models being favourable with respect to the Planck 2018 data, it is clear from comparing the two solutions, such as in figure (\ref{Model_Comp}), that the two solutions deviate quite dramatically long before either model enters the region of parameter space favoured by the Planck 2018 data. This is because the higher order terms actually become large during inflation and are therefore relevant to the calculation of observables and dramatically change the inflationary predictions of the model by shifting the $r-n_s$ spectrum far to the left. Clearly the QH and QHS models need to be treated distinctly in the context of inflation, and that one cannot simply take the Quartic Hilltop model as an accurate approximation of Quartic Hilltop Squared.
    
    Throughout this work frequent references and comparisons are made to numerical solutions for the Hilltop models in question. The numerical solutions help evaluate the validity of the results and form part of ongoing work by the author to develop numerical solutions for a wide range of inflationary models. The ability to solve models numerically will aid in investigating the generic features of Hilltop models. For example whether or not the $N_{\rm end} \propto \lambda^{-1/2}$ relationship holds for other Hilltop models or if there is some other power law. Furthermore, the numerical solutions also make it possible to investigate how dramatically the $r-n_s$ changes in other Hilltop models, say for example quintic models as opposed to quartic and whether or not those models require us to account for the higher order correction terms.
    
   At a first glance, the fact that both the QH and QHS model predictions for the tensor-scalar ratio and spectral index fit nicely into the Planck 2018 bounds seems encouraging, however as we note, and has already been discussed at length in the literature, this does not necessarily mean the model goes a long way in helping explain the data that has been presented since we may generate any $(n_s,r)$ pair by appropriate choice of $N_k$, that may or may not correspond to reheating predictions that are consistent with standard cosmology.

    Correcting the models UV behaviour allows us to explore the reheating dynamics of the QHS model through the numerical solutions and further constrain its $r-n_s$ parameter space to the region bounded by $63 \leq N_k \leq 68$, by considering Big Bang Nucleosynthesis limits on the reheating temperature. Whilst this does not drastically reduce the range of allowed values of $n_s$ and $r$, the area of $(n_s,r)$ pairs is significantly smaller than if one were to just consider the bounds derived by the Planck 2018 survey.

\section*{Acknowledgements}
The authors are indebted to the anonymous referee whose input has greatly improved this paper.

\bibliography{refs.bib}

\bibliographystyle{ieeetr}

\end{document}